# *Melting behavior and dynamical properties of $Cr_2Ge_2Te_6$ phase-change material*

Suyang Sun[1,2], Yihui Jiang[2], Riccardo Mazzarello[3], Wei Zhang[2,*]

[1]Institute of Materials, Henan Academy of Sciences, Zhengzhou, 450046, China.
[2]Center for Alloy Innovation and Design (CAID), State Key Laboratory for Mechanical Behavior of Materials, Xi'an Jiaotong University, Xi'an, 710049, China.
[3]Department of Physics, Sapienza University of Rome, Rome00185, Italy

*Corresponding author: wzhang0@mail.xjtu.edu.cn

**Abstract**
$Cr_2Ge_2Te_6$ (CrGT) is known as an intrinsic ferromagnetic semiconductor and a promising candidate for phase-change memory applications. In amorphous CrGT, Cr atoms form non-defective octahedral motifs with Te atoms, similar to those in the crystalline phase. The abundance of $Cr[Te_6]$ octahedra is regarded as the key structural factor in reducing the resistance drift coefficient of amorphous CrGT. However, the stage at which these octahedra emerge during melt-quench amorphization remains unclear. Here, we present *ab initio* molecular dynamics (AIMD) simulations to model the melting process of crystalline CrGT and to investigate the dynamical properties of liquid and supercooled liquid CrGT in detail. Upon heating, Ge atoms are observed to leave their lattice sites earlier than Cr and Te atoms, diffusing into the van der Waals gap and initiating the collapse of the layered structure. The $Cr[Te_6]$ octahedra are more robust, maintaining their structural pattern up to 1400 K despite continuous rupture and re-formation of Cr-Te bonds. At higher temperatures, Cr and Te atoms start to migrate independently. In supercooled liquid CrGT at 550 K, most Cr-centered octahedra remain intact, with only limited Cr-Te bond breaking. The collective motion of these octahedra in this temperature regime helps explain why crystallization in CrGT devices can be accomplished in tens of nanoseconds.

Chalcogenide phase-change materials (PCM) are among the leading candidates for non-volatile memory,[1-3] neuromorphic computing[4-6] and reconfigurable nanophotonics[7-9]. The basic working principle relies on the large variation in electric and optical properties upon phase transition between amorphous and crystalline phases, which is exploited to encode digital information.[1] The most established PCMs are found in the Ge-Sb-Te ternary diagram, which comprises three major materials families, namely, the pseudo-binary $(GeTe)_m(Sb_2Te_3)_n$ alloys,[10-12] Sb-Te binary alloys [13-15] and doped Sb alloys [16-18] with three-dimensional (3D) crystalline structures. The crystallization temperature $T_x$ of the flagship $Ge_2Sb_2Te_5$ alloy (GST) is ~150 °C,[11] and such amorphous stability is already sufficient for stand-alone memory devices, which operate in the temperature range of 60-80 °C. However, embedded memory devices require higher amorphous stability. Alloying GST with excess Ge or C atoms largely increases the crystallization temperature, and these non-stoichiometric PCMs have been commercialized recently.[19-21] Nevertheless, the pursuit of PCMs with intrinsically high $T_x$ continues, as such materials are expected to exhibit reduced phase segregation issues upon extensive cycling.

Crystalline (c-) $Cr_2Ge_2Te_6$ (CrGT) is a two-dimensional (2D) layered van der Waals (vdW) semiconductor that shows intrinsic ferromagnetism[22] even when its thickness is reduced to a single atomic slab.[23] Amorphous (a-) CrGT thin films exhibit a high $T_x$ of ~276 °C,[24-28] and their resistivity increases by nearly two orders of magnitude upon crystallization.[25] The reversible phase transition between the two solid state phases can be achieved by applying electrical pulses with duration of tens to hundreds of nanoseconds. In addition, a-CrGT shows a smaller drift coefficient [25] than a-GST during spontaneous glass relaxation,[29] enabling more reliable information retention over time. These features make CrGT a promising candidate for embedded memory applications. Moreover, CrGT preserves its spin-polarized nature in the amorphous phase and forms a spin glass state below 20 K.[30] The tunable magnetic properties of CrGT [22,30] can be exploited for phase-change magnetic switching applications.[31]

Recently, we carried out density functional theory (DFT) based *ab initio* molecular dynamics (AIMD) simulations to investigate the structural and electronic properties of amorphous CrGT.[30] We observed a major change in the local bonding environment of Ge atoms, shifting from tetrahedral to defective octahedral motifs upon amorphization, whereas Cr atoms maintained similar octahedral motifs with Te atoms in both phases. Spin polarization was found to be the key factor stabilizing these Cr-centered octahedra in the amorphous phase. Notably, the abundance of intact $Cr[Te_6]$ units was revealed to be the origin of no-drift behavior in the closely related amorphous alloy $CrTe_3$.[32] However, it remains elusive at which stage of the melt-quenched amorphization process such octahedral fragments first emerge. In the following, we focus on the melting behavior of CrGT and the dynamical properties of the disordered CrGT phase at elevated temperatures, aiming to achieve an in-depth understanding of the amorphization process.

We carried out AIMD simulations using the CP2K package.[33] Goedecker pseudopotentials,[34] the Perdew-Burke-Ernzerhof (PBE) functional [35] and semi-empirical van der Waals corrections [36] were employed. The Kohn–Sham orbitals were expanded in a Gaussian-type basis set with double-/triple-ζ polarization quality, whereas the charge density was expanded in plane waves, with a cutoff of 300 Ry. The AIMD calculations were performed using the second-generation Car-Parrinello scheme [37] in the canonical (NVT) ensemble with a time step of 2 fs. Spin polarization was included by treating the $\alpha$ and $\beta$ spin channels independently, without imposing any spin-symmetry constraints. All the AIMD calculations were carried out within a spin-polarized framework.

Figure 1a shows the crystalline phase of CrGT, which adopts a hexagonal structure with three atomic slabs stacked along the c-axis (space group $P\bar{3}m1$). Every Ge atom (blue) forms a tetrahedral motif with three Te atoms (green) and one Ge atom, and every Cr atom (red) forms an octahedral motif with six Te atoms. In an atomic slab of the hexagonal unit cell, two Cr[Te$_6$] corner-shared octahedra and two intergrown Ge[GeTe$_3$] tetrahedra share the same set of Te atoms. To simulate the melting process, we constructed a 3×4×1 orthorhombic supercell, containing 72 Cr atoms, 72 Ge atoms and 216 Te atoms. The lattice edges of this supercell were 20.65 Å, 23.84 Å and 20.36 Å, respectively. This supercell model was then heated from 300 K to 1700 K in 180 ps. Figure 1b shows the heating process with a heating rate of ~8 K/ps, together with the evolution of the total energy as a function of time. The kinetic energy scales with the preset temperatures, while a rapid increase in the potential energy is observed between 126 and 135 ps, during with the average temperature is 1345±67 K. Afterwards, the potential energy again scales with the target temperatures.

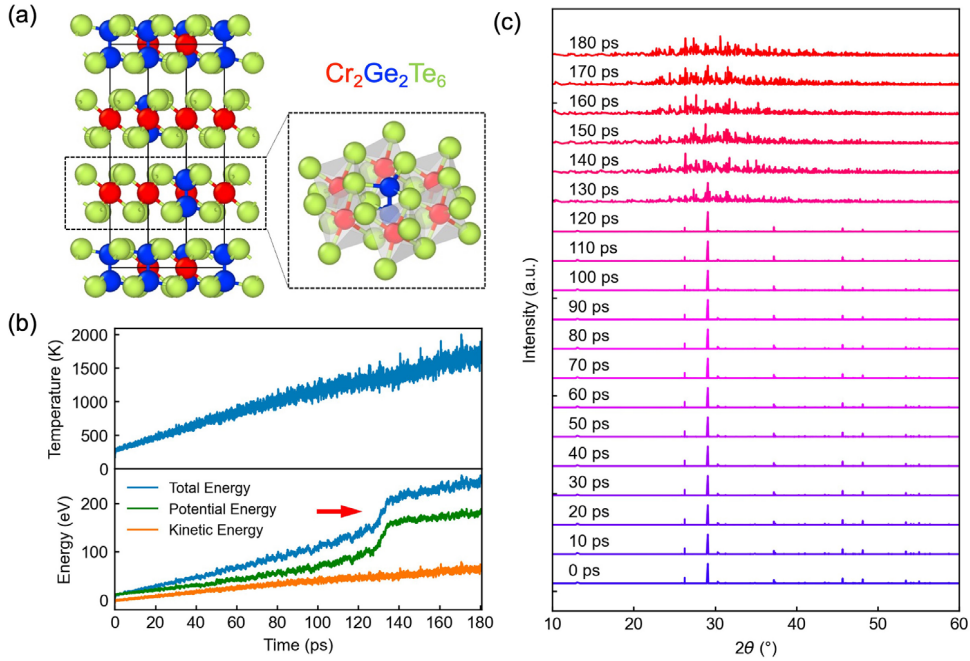

**Figure 1.** AIMD simulations of the melting process. (a) Structural model of crystalline CrGT. (b) AIMD simulations of c-CrGT during heating. The red arrow indictaes the melting transition. (c) Simulated XRD patterns at different stages of the heating process.

Multiple snapshots were extracted from the heating trajectory, and their X-ray diffraction (XRD) patterns were simulated using Cu K$_\alpha$ radiation with the pymatgen package.[38] As seen in Figure 1c, the initial crystalline model exhibits prominent peaks at $2\theta = $ ~29.1° and ~26.2°, corresponding to the (113) and (006) Bragg reflections, in agreement with the experimental XRD pattern.[22] This structural pattern is preserved up to 120 ps, beyond which the characteristic diffraction peaks vanish. All the structural snapshots used for the XRD simulations are provided in Figure S1. Both the energy profiles and the XRD simulations consistently indicate a clear melting transition at around 130 ps during the heating process.

To gain further understanding of the melting process, we analyzed the evolution of atomic coordinates along the *c* direction. Figure 2a displays the initial atomic configuration of the orthorhombic supercell model, and Figure 2b records the variations in the atomic positions of the Cr, Ge and Te atoms. The shaded gray regions indicate the vdW gaps of the initial crystalline structure (~3.36 Å). Overall, Ge atoms are more sensitive to high temperatures than Cr and Te atoms. From around 50 ps, all Ge atoms exhibited strong vibrations around their lattice positions, and some started to reach the edges of the vdW regions. At 100 ps, when the average temperature just exceeded 1000 K, Ge atoms already spread across both the atomic-slab regions and the vdW gaps. In contrast, Cr and Te atoms maintained their positions until approximately 130 ps, after which the 2D layered features collapsed completely. Figure 2c shows three representative snapshots: at 60 ps, arrows indicate the Ge atoms deviating from their lattice positions; at 110 ps, dashed circles highlight the Ge atoms that left the atomic-slab regions and entered the vdW gaps; and at 180 ps, the model was fully melted.

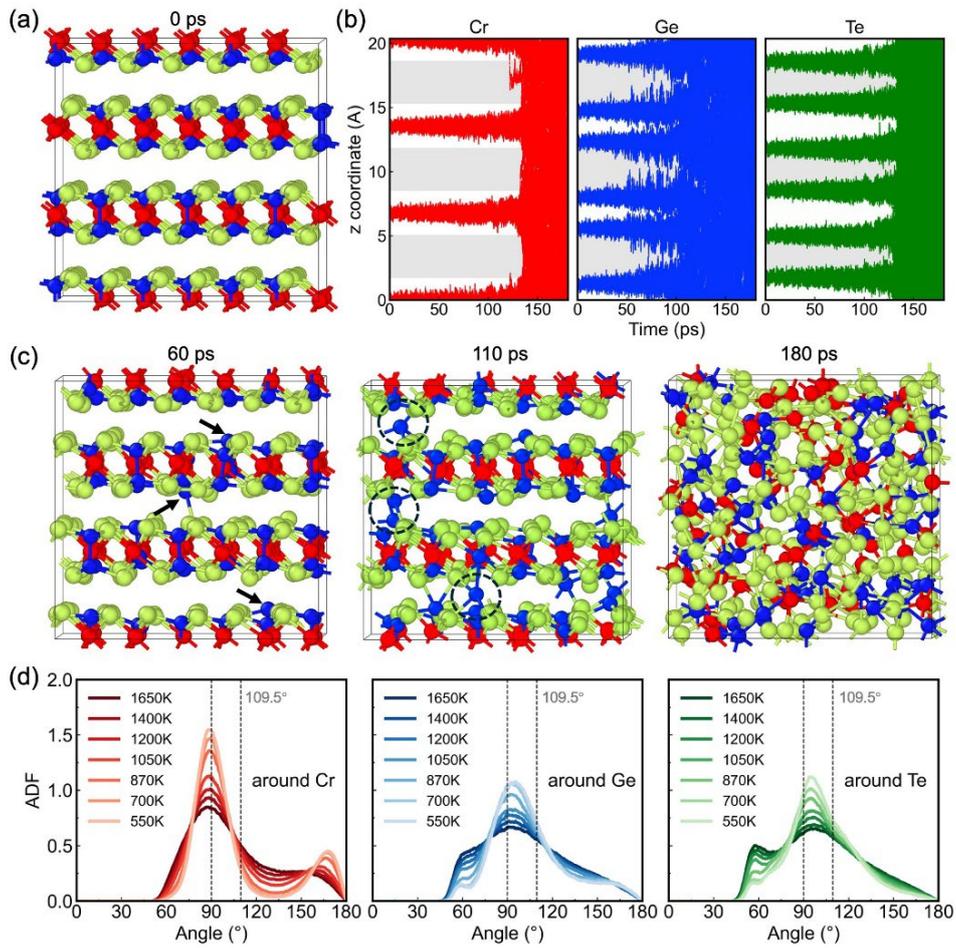

**Figure 2.** Atomic motions upon heating. (a) Initial supercell model of the ordered crystalline phase, in which Cr, Ge and Te atoms are rendered by red, blue and green spheres, respectively. (b) Element-resolved evolution of atomic coordinates along the *c* direction. (c) Snapshots of the CrGT model at 60 ps, 110 ps, and 180 ps, respectively. (d) ADF curves around Cr, Ge and Te atoms as a function of temperature. The dashed vertical lines mark the characteristic angles of 90° and 109.5°, corresponding to ideal octahedral and tetrahedral geometries, respectively. The dynamical cutoffs used, Cr-Cr 2.51, Cr-Ge 2.80, Cr-Te 2.99, Ge-Ge 3.10, Ge-Te 3.18 and Te-Te 3.13 Å, were taken from our previous work, in which they were derived from the chemical bonding analysis of a-CrGT [30].

We also generated an additional model of liquid CrGT by heating a quasi-random structure (constructed using a random number generator) at 1650 K for 50 ps. Figure S2 compares the radial distribution functions (RDFs) of this model with those of the previous model. The small differences in both the total and partial RDFs confirm that the crystalline model shown in Figure 2 was fully melted after 180 ps. This liquid model was subsequently quenched to the experimentally measured $T_x$, ~550 K, at a rate of ~10 K/ps. Several snapshots were taken at each temperature for independent AIMD simulations of 50 ps, and the final 40 ps of each trajectory were used for the analyses of structural and dynamical properties.

According to the angular distribution function (ADF) analysis (Figure 2d), both Cr and Ge atoms tend to form octahedral bonds with Te atoms at elevated temperatures. Even at 1650 K, at which the atoms are only transiently connected, the bond angles around Cr atoms are predominantly near 90°. As the temperature decreases, the ADF values at other angles drop sharply. At 550K, the bond angle distribution around Cr atoms shows a dominant peak at 90° and a secondary peak near 180°, consistent with the structural features of a-CrGT at 300 K. We also calculated the element-resolved mean squared displacement (MSD) of CrGT following the equation (1),

$$MSD = \langle R^2(t) \rangle = \frac{1}{N}\langle \sum_{i=1}^{N} |\boldsymbol{R}_i(t+t_0) - \boldsymbol{R}_i(t_0)|^2 \rangle \quad (1)$$

where $\boldsymbol{R}_i$ denotes the position of atom $i$ of the specific element, $t_0$ is the initial point in time, $N$ is the number of atoms of the specific element. Based on the MSD results, the diffusion coefficient $D$ values were calculated by $D = \frac{1}{6}\frac{\partial}{\partial t}\lim_{t \to \infty}\langle R^2(t) \rangle$.

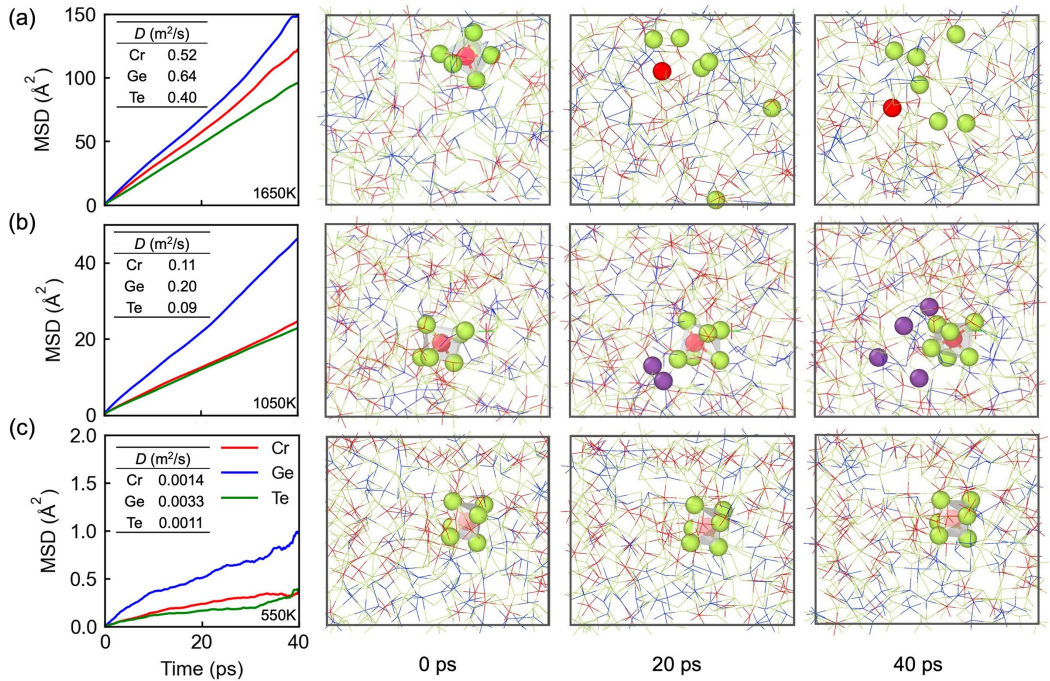

**Figure 3**. Element-resolved MSD curves and MD snapshots of CrGT at (a) 1650K, (b) 1050K and (c) 550K. A Cr[Te$_6$] octahedron was highlighted and tracked over time at each temperature. The purple spheres are the Te atoms that migrated away from the original Cr[Te$_6$] octahedron. The dynamical cutoff for Cr-Te is increased to 3.4 Å for better visualization of the octahedral motif.

The MSD profiles of Cr, Ge and Te atoms are shown in Figure 3 and Figure S3, together with representative MD snapshots taken at different temperatures. The extracted total and element-resolved diffusion coefficients $D$ are listed in Table S1. At all temperatures, $D_{Ge}$ is consistently much larger than $D_{Cr}$ and $D_{Te}$. The CrGT model shows typical liquid-like behavior[39-41] at 1650 K, where the Ge, Cr and Te atoms migrate nearly independently, with coefficients $D_{Ge} \sim 0.64$ m$^2$/s, $D_{Cr} \sim 0.52$ m$^2$/s and $D_{Te} \sim 0.40$ m$^2$/s. Although Cr[Te$_6$] octahedra can form transiently, the Cr atom and its six coordinating Te atoms quickly decouple over time, as highlighted by the polyhedron in Figure 3a. Below 1400 K, the small differences between the MSD curves of Cr and Te suggest a correlation between their atomic motion during diffusion. As shown in Figure 3b, the initial Cr[Te$_6$] octahedron gradually loses Te atoms (colored in purple) but rapidly incorporates nearby Te atoms (colored in green), thereby preserving the octahedral motif at 1050 K. The four purple Te atoms become bonded with other Cr atoms to form new Cr-centered octahedra. In the temperature range between 1400 K and 870 K, Cr-centered octahedra are continuously observed, accompanied by frequent rupture and re-formation of Cr-Te octahedral bonds. When the temperature is further reduced to 550-700 K, the atomic mobility decreases sharply and the MSD curves of Cr and Te atoms show caging effects. The total $D$ value of supercooled liquid CrGT is ~0.0016 m$^2$/s at 550 K, about 300 times smaller than its value at 1650 K. Cr-centered octahedra move collectively with only limited rupture of Cr-Te bonds. One such octahedron is highlighted in Figure 3c.

Based on the diffusion coefficient data from 1650 K to 550 K, we calculated the activation energy of diffusion for each element following the Arrhenius equation,

$$\ln D = -\frac{E_a}{k_B}\frac{1}{T} + \ln D_0 \qquad (2)$$

where $k_B$ is the Boltzmann constant, $D_0$ is a pre-exponential factor, and $E_a$ denotes the activation energy. The computed logarithmic $D$ over 1000/T profiles for Cr (red), Ge (blue) and Te atoms (green) are shown in Figure 4a, with the fitted slopes of $-E_a/k_B$. Within the simulated temperature range, the $E_a$ of Cr and Te atoms are quite similar (approximately 0.44 eV) and are both larger than that of Ge atoms (0.37 eV). The higher diffusion activation energies of Cr and Te compared to Ge suggest that these atoms have to overcome higher energy barriers during atomic migrations.

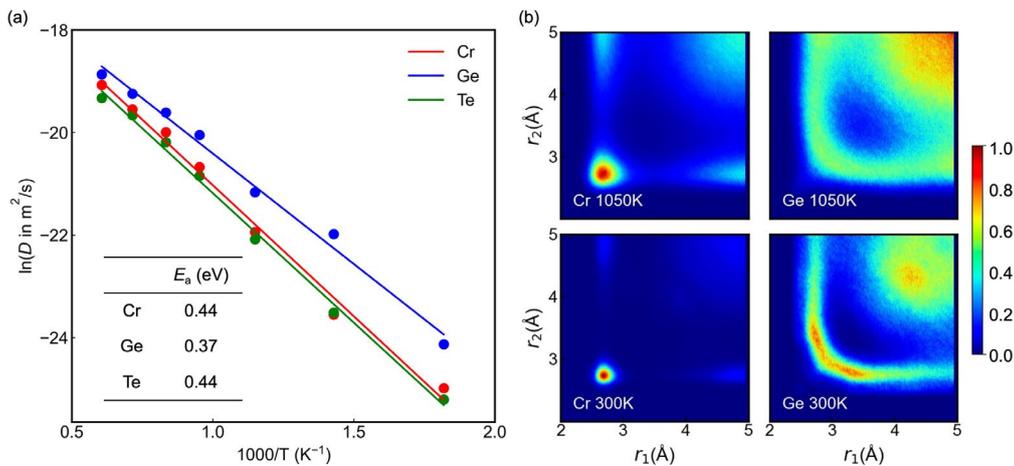

**Figure 4.** Dynamical and structural features of disordered CrGT. (a) Element-resolved activation energy $E_a$ for diffusion, obtained from Arrhenius fits. (b) ALTBC analysis of bond-length distributions for Cr- and Ge-centered directional bonding pairs in disordered CrGT at 1050 K and 300 K, respectively.

The collection motions of Cr[Te$_6$] octahedra in the supercooled liquid state at 550-700 K may explain why crystallization of CrGT could be completed using a 30 ns electrical pulse in devices.[24] In conventional PCMs, crystal growth at the amorphous-to-crystal interfaces proceeds via frequent impinging and statistical sticking of individual atoms. However, for CrGT, when one Te atom is sticked to the crystal forefront, the associated Cr[Te$_6$] of this Te atom could be captured by this crystalline domain. Nevertheless, the activation barrier for the diffusion of the Cr-centered octahedra is higher than that for individual atoms. Besides, the incubation of a CrGT seed also requires a major change in bonding configuration of Ge-centered units from defective octahedral motifs (Figure 2d) to homopolar Ge-Ge bonds mediated intergrown Ge[GeTe$_3$] tetrahedral motifs (Figure 1a). These two features suggest that it is unlikely that the crystallization time can be further reduced to a few ns or sub-nanosecond[42,43] for CrGT devices. The formation of complex 2D seeds results in a high nucleation barrier for CrGT, making it a growth-type PCM with high $T_x$.[44]

The above analyses conclude that the Cr-centered octahedra already start to form at 1400 K, and the probability of Cr-Te bond rupture is largely reduced at 550 K. At even lower temperatures, Cr-centered octahedra should remain intact with time. In the temperature range between 300 K and 450 K, glass relaxation of conventional a-PCMs typically cause a steady increase in electrical resistance with time, making accurate multilevel programming difficult.[29] The a-CrGT thin film shows a smaller drift coefficient ($v$~0.03) than that of a-GST ($v$~0.11) at 373 K.[32] Figure 4b displays the angular-limited three-body correlation (ALTBC) function[45] for Cr and Ge atoms, which shows the distribution of short and long bonds among atomic pairs with bond angles close to 180°. At both 1050 K and 300 K, Cr-centered chemical bonds show no visible Peierls distortion with one major peak. Regarding Ge-centered units, the degree of Peierls distortion is much enhanced upon cooling, resulting in an extended double-wings shape ALTBC at room temperature. Such bonding feature is consistent with conventional PCMs.[46] Therefore, the reinforcement of Peierls distortion around Ge atoms[45] should contribute to the resistance drift of a-CrGT. When Ge atoms are all removed, the amorphous phase of CrTe$_3$ is only consisted of Cr[Te$_6$] octahedra, and there is no tendency for Peierls distortion within each and every octahedron. Indeed, a-CrTe$_3$ thin films and devices show practically no resistance drift across a wide temperature range.[32]

In conclusion, we have investigated the atomic-scale details of the melting process and the dynamical properties of liquid and supercooled liquid CrGT via spin-polarized AIMD simulations. Upon heating, Ge atoms are more sensitive to thermal excitation than Cr and Te, becoming readily displaced from their equilibrium sites and entering the vdW gaps well before the collapse of Cr[Te$_6$] octahedra. The dynamical behavior of CrGT can be categorized into three regimes: (1) conventional atomic diffusion above 1650 K; (2) individual Ge diffusion accompanied by correlated Cr/Te diffusion between 870 K and 1400 K and (3) individual Ge diffusion combined with collective motions of Cr-centered octahedra between 550 K and 700 K, the relevant temperature range for fast crystallization. These regimes correspond to high, medium, and low Cr–Te bond-rupture frequencies, respectively. Given the limited statistical sampling and the challenge in defining chemical bonds at high temperatures, the boundaries between these regimes should be considered as approximate.

Our work shows that the formation of Cr-centered octahedra already starts near 1400 K and becomes substantially stabilized at 550 K and below. The robustness and abundance of Cr[Te$_6$] octahedra likely reduce structural aging in a-CrGT, resulting in a low drift coefficient. Regarding crystallization, the presence of Cr[Te$_6$] octahedra could promote efficient fragment-based crystal growth, while the complex two-dimensional arrangement of Cr[Te$_6$] octahedra and Ge[GeTe$_3$] tetrahedra sharing common Te atoms sets a large nucleation barrier. As a result, CrGT simultaneously exhibits a high crystallization temperature of 276 ºC and fast crystallization speed of 30 ns under electrical programming. These insights help elucidate the structural and dynamical differences between 2D CrGT/CrTe$_3$ PCMs and conventional 3D PCMs at the microscopic level.

**Competing interests**
The authors declare no competing interests.

**Data Availability Statement**
The data that support the findings of this study will be available in the CAID Repository at https://caid.xjtu.edu.cn/info/1003/2225.htm, reference number 2225.

**Acknowledgments**
W. Z. acknowledges the support of the National Natural Science Foundation of China (62374131) and the 111 Plan (B25007). S. S. thanks the Fundamental Research Fund of Henan Academy of Sciences (project no. 251817088). R.M. gratefully acknowledges funding from the PRIN 2020 project "Neuromorphic devices based on chalcogenide heterostructures" funded by the Italian Ministry for University and Research (MUR). We acknowledge the HPC platform of XJTU and Computing Center in Xi'an for providing computational resources. The International Joint Laboratory for Micro/Nano Manufacturing and Measurement Technologies of XJTU is acknowledged.

Supporting Information

for

*Melting behavior and dynamical properties of Cr$_2$Ge$_2$Te$_6$ phase-change material*

Suyang Sun[1,2], Yihui Jiang[2], Riccardo Mazzarello[3], Wei Zhang[2,*]

[1]Institute of Materials, Henan Academy of Sciences, Zhengzhou, 450046, China.
[2]Center for Alloy Innovation and Design (CAID), State Key Laboratory for Mechanical Behavior of Materials, Xi'an Jiaotong University, Xi'an, 710049, China.
[3]Department of Physics, Sapienza University of Rome, Rome00185, Italy
*Corresponding author: wzhang0@mail.xjtu.edu.cn

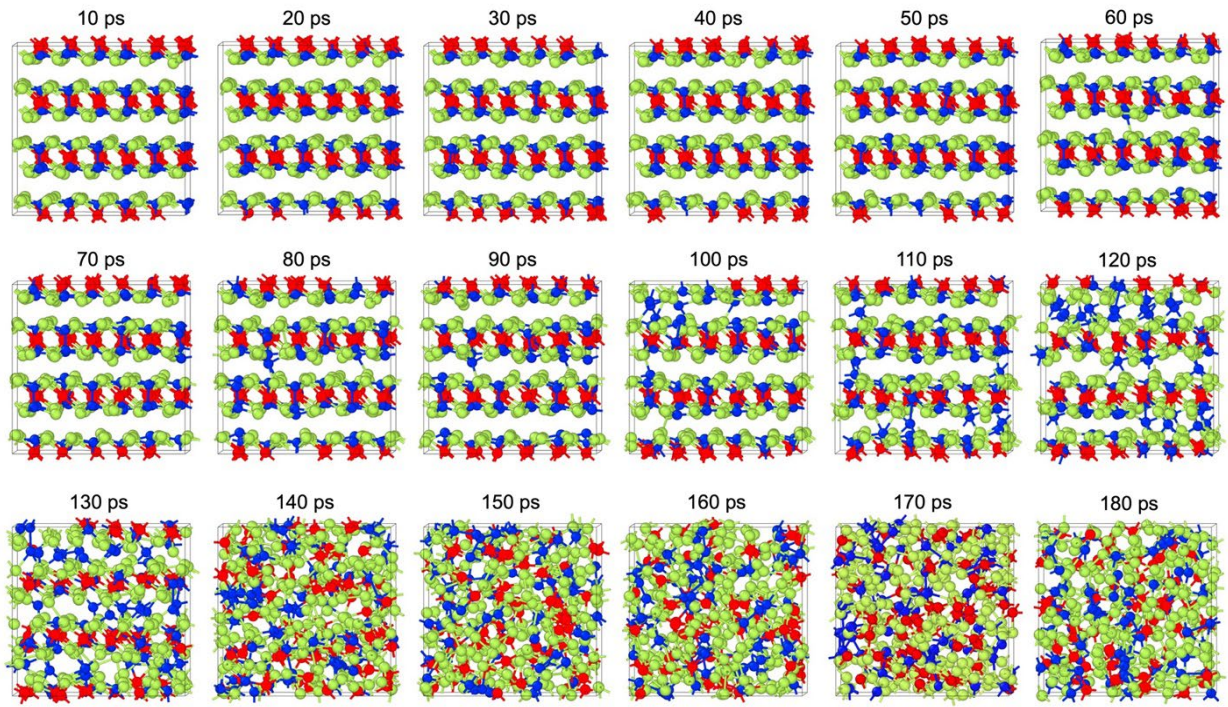

**Figure S1**. The MD snapshots used for XRD pattern simulations shown in Figure 1c.

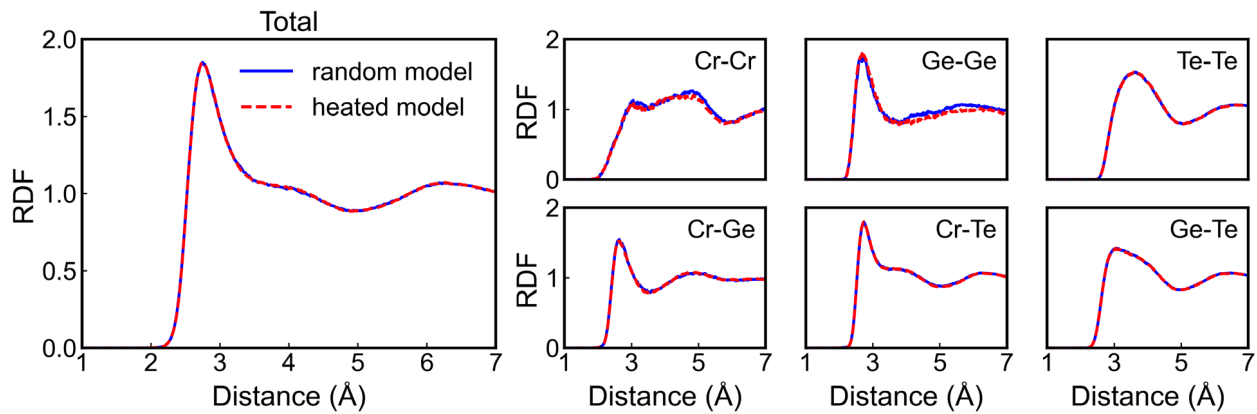

**Figure S2.** The total and partial RDFs of the melted model and the liquid model started from a fully randomized model (random model) at ~1650 K.

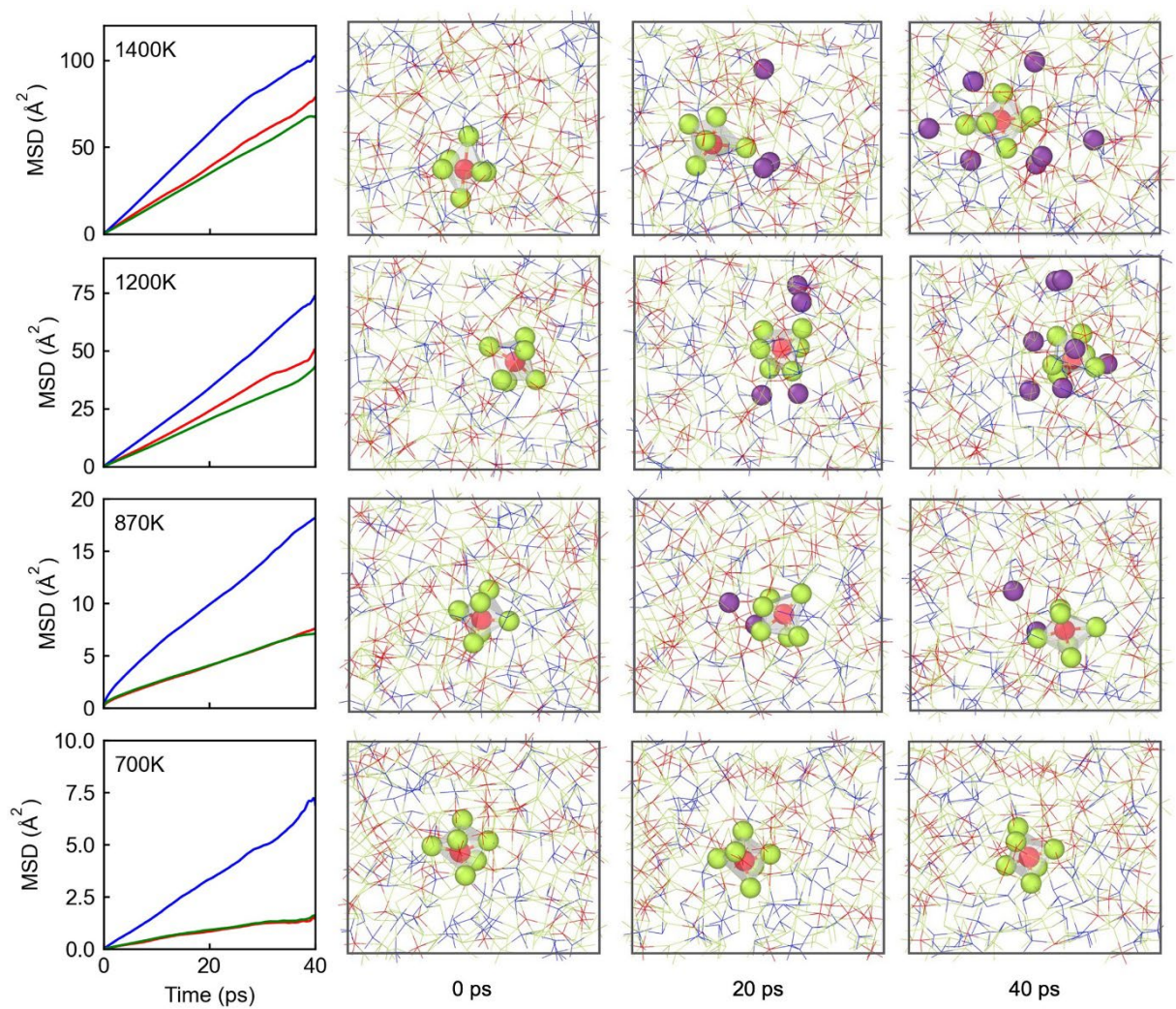

**Figure S3.** The element-resolved MSD curves and MD snapshots of $Cr_2Ge_2Te_6$ at 1400 K, 1200 K, 870 K and 700 K, respectively. The purple spheres are the Te atoms that migrated away from the original $Cr[Te_6]$ octahedron. The dynamical cutoff for Cr-Te is enlarged to 3.4 Å for better visualization of the octahedral motif.

| Temperature (K) | Diffusion coefficient $D$ (m$^2$/s) | | | |
| --- | --- | --- | --- | --- |
| | Total | Cr | Ge | Te |
| 1650 | 0.48 | 0.52 | 0.64 | 0.40 |
| 1400 | 0.33 | 0.32 | 0.44 | 0.29 |
| 1200 | 0.21 | 0.21 | 0.31 | 0.17 |
| 1050 | 0.11 | 0.11 | 0.20 | 0.09 |
| 870 | $0.34 \times 10^{-1}$ | $0.30 \times 10^{-1}$ | $0.65 \times 10^{-1}$ | $0.26 \times 10^{-1}$ |
| 700 | $0.11 \times 10^{-1}$ | $0.60 \times 10^{-2}$ | $0.28 \times 10^{-1}$ | $0.62 \times 10^{-2}$ |
| 550 | $0.16 \times 10^{-2}$ | $0.14 \times 10^{-2}$ | $0.33 \times 10^{-2}$ | $0.11 \times 10^{-2}$ |

**Table S1**. The diffusion coefficients $D$ calculated based on the MSD curves shown in Figure 3 and Figure S3.